\documentclass[twocolumn,showpacs,preprintnumbers,amsmath,amssymb]{revtex4}

\usepackage{graphicx}
\usepackage{dcolumn}
\usepackage{bm}
\usepackage{epsfig,amssymb}

\begin{document}


\title{Vortex proliferation in the Berezinskii-Kosterlitz-Thouless regime on a two-dimensional lattice of Bose-Einstein condensates}

\author{V. Schweikhard, S. Tung, and E.~A. Cornell\cite{qpdNIST}}
\affiliation{JILA, National Institute of Standards and Technology and University of Colorado, and Department of Physics, University of Colorado,
Boulder, Colorado 80309-0440}
\date{\today}

\begin{abstract}
We observe the proliferation of vortices in the Berezinskii-Kosterlitz-Thouless regime on a two-dimensional array of Josephson-coupled
Bose-Einstein condensates. As long as the Josephson (tunneling) energy $J$ exceeds the thermal energy $T$, the array is vortex-free. With
decreasing $J/T$, vortices appear in the system in ever greater numbers. We confirm thermal activation as the vortex formation mechanism and
obtain information on the size of bound vortex pairs as $J/T$ is varied.
\end{abstract}

\maketitle

One of the defining characteristics of superfluids is long-range phase coherence \cite{Leggett}, which may be destroyed by quantum fluctuations,
as in the Mott-insulator transition \cite{GreinerMI,ZollerMI}, or thermal fluctuations, e.g. in one-dimensional Bose gases
\cite{1dfluctuationsA,1dfluctuationsB} and in a double-well system \cite{Oberthaler}. In two dimensions (2D), Berezinskii \cite{B}, Kosterlitz
and Thouless \cite{KT} (BKT) developed an elegant description of thermal phase fluctuations based on the unbinding of vortex-antivortex pairs,
i.e. pairs of vortices of opposite circulation. The BKT picture applies to a wide variety of 2D systems, among them Josephson junction arrays
(JJA), i.e. arrays of superfluids in which phase coherence is mediated via a tunnel coupling $J$ between adjacent sites. Placing an isolated
({\it free}) vortex into a JJA is thermodynamically favored if its free energy $F=E-TS\leq0$. In an array of period $d$ the vortex energy
diverges with array size $R$ as $E\approx J\log(R/d)$ \cite{Tinkham}, but may be offset by an entropy gain $S\approx \log(R/d)$ due to the
available $\approx R^2/d^2$ sites. This leads to a critical condition $(J/T)_{crit}\approx1$ independent of system size, below which free
vortices will proliferate. In contrast, {\it tightly bound} vortex-antivortex {\it pairs} are less energetically costly and show up even above
$(J/T)_{crit}$. The overall vortex density is thus expected to grow smoothly with decreasing $J/T$ in the BKT crossover regime.

\begin{figure}
\epsfig{figure=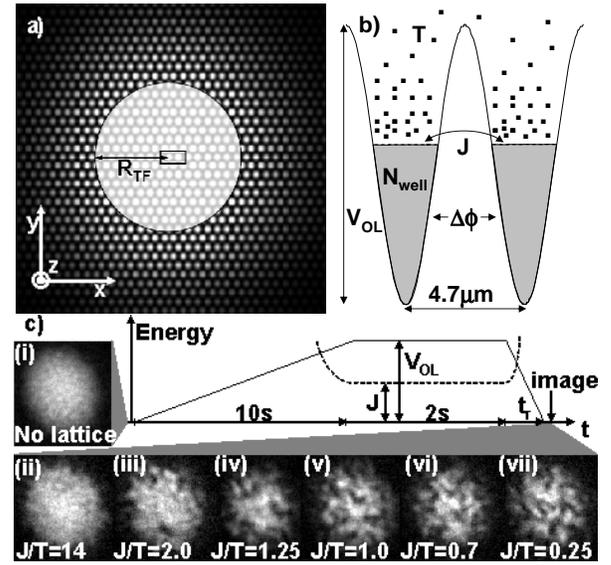,width=0.9\linewidth,clip=} \caption{\label{Fig1} Experimental system. (a) 2D optical lattice intensity
profile. A lattice of Josephson-coupled BECs is created in the white-shaded area. The central box marks the basic building block of our system,
the double-well potential shown in (b). The barrier height $V_{OL}$ and the number of condensed atoms per well, $N_{well}$, control the
Josephson coupling $J$, which acts to lock the relative phase $\Delta\phi$. A cloud of uncondensed atoms at temperature T induces thermal
fluctuations and phase defects in the array when $J<T$. (c) Experimental sequence: A BEC (i) is loaded into the optical lattice over $10\,s$,
suppressing $J$ to values around T. We allow $2\,s$ for thermalization. To probe the system, we ramp off the lattice on a faster timescale $t_r$
\cite{rampdown} and take images of the recombined condensate. When $J$ is reduced below T (ii)-(vii), vortices (dark spots) appear as remnants
of the thermal fluctuations in the array.}
\end{figure}

Transport measurements, both in continuous superfluids \cite{Reppy89,Safonov} and superconducting JJA \cite{Resnick} have confirmed the
predictions of BKT, without however directly observing its microscopic mechanism, vortex-antivortex unbinding. A recent beautiful experiment
\cite{Hadzibabic} in a continuous 2D Bose gas measured the phase-phase decay function through the BKT cross-over, and saw evidence for thermal
vortex formation. For related theoretical studies see e.g. \cite{Simula}. In this work we present more detailed vortex-formation data, collected
in a 2D array of BECs with experimentally controllable Joephson couplings. The system was studied theoretically in \cite{Trombettoni}.
\par
Our experiment starts with production of a partially Bose-condensed sample of $^{87}$Rb atoms in a harmonic, axially symmetric magnetic trap
with oscillation frequencies $\{\omega_{x,y},\omega_{z}\}=2\pi\{6.95,15.0\}\,\rm{Hz}$. The number of {\it condensed} atoms is kept fixed around
$6\times10^5$ as the temperature is varied. We then transform this system into a Josephson junction array, as illustrated in Fig. \ref{Fig1}. In
a $10\,s$ linear ramp, we raise the intensity of a 2D hexagonal optical lattice \cite{LatticeConfig} of period $d=4.7\mu m$ in the x-y plane.
The resulting potential barriers of height $V_{OL}$ between adjacent sites [Fig.\ref{Fig1}(b)] rise above the condensate's chemical potential
around $V_{OL}\approx 250-300\,Hz$, splitting it into an array of condensates which now communicate only through tunneling. This procedure is
adiabatic even with respect to the longest-wavelength phonon modes of the array \cite{Javanainen,Burnett} over the full range of $V_{OL}$ in our
experiments. Each of the $\approx190$ occupied sites (15 sites across the BEC diameter $2\times R_{TF}\approx 68\,\mu m$ \cite{RadialEffects})
now contains a macroscopic BEC, with $N_{well}\approx7000$ condensed atoms in each of the central wells at a temperature $T$ that can be
adjusted between $30-70\,nK$. By varying $V_{OL}$ in a range between $500\,Hz$ and $2\,kHz$ we tune $J$ between $1.5\,\mu K$ and $5\,nK$,
whereas the ``charging'' energy $E_c$, defined in \cite{Leggett}, is on the order of a few $pK$, much smaller than both $J$ and $T$. In this
regime, thermal fluctuations of the relative phases $\Delta\phi_{Th}\approx\sqrt{T/J}$ are expected, while quantum fluctuations
$\Delta\phi_{Q}\approx(E_c/4J)^{1/4}$ are negligible \cite{Leggett}.
\par
The suppression of the Josephson coupling greatly suppresses the energy cost of phase fluctuations in the x-y plane, {\it between} condensates,
$J[1-\cos(\Delta \phi)]$, compared to the cost of axial (z) phase fluctuations {\it inside} the condensates \cite{Walraven}. As a result, axial
phase fluctuations remain relatively small, and each condensate can be approximated as a single-phase object \cite{Axialphase}.
\par
After allowing $2\,s$ for thermalization, we initiate our probe sequence. We first take a nondestructive thermometry image in the x-z plane,
from which the temperature $T$ and, from the axial condensate size $R_z$, the number of condensed particles per well, $N_{well}$, is obtained
(see below). To observe the phase fluctuations we then turn down the optical lattice on a time-scale $t_r$ \cite{rampdown}, which is fast enough
to trap phase winding defects, but slow enough to allow neighboring condensates to merge, provided their phase difference is small. Phase
fluctuations are thus converted to vortices in the reconnected condensate, as has been observed in the experiments of Scherer {\it et al.}
\cite{Anderson}. We then expand the condensate by a factor of 6 and take a destructive image in the x-y plane.
\par
Figure \ref{Fig1}(c) illustrates our observations: (ii)-(vii) is a sequence of images at successively smaller $J/T$ (measured in the center of
the array \cite{Javeragenote}). Vortices, with their cores visible as dark ``spots'' in (iii)-(vii), occur in the BEC center around $J/T=1$.
Vortices at the BEC edge appear earlier, as here the magnetic trap potential adds to the tunnel barrier, suppressing the {\it local} $J/T$ below
the quoted value. That the observed ``spots'' are indeed circulation-carrying vortices and antivortices is inferred from their slow $\approx
100\,ms$ decay after the optical lattice ramp-down, presumably dominated by vortex-antivortex annihilation. From extensive experiments on
vortices in our system we know that circulation-free ``holes'' fill so quickly due to positive mean field pressure, that they do not survive the
pre-imaging expansion. Vortices with identical circulation would decay by dissipative motion to the BEC edge, in our trap over $\gtrsim10\,s$.

\begin{figure}
\epsfig{figure=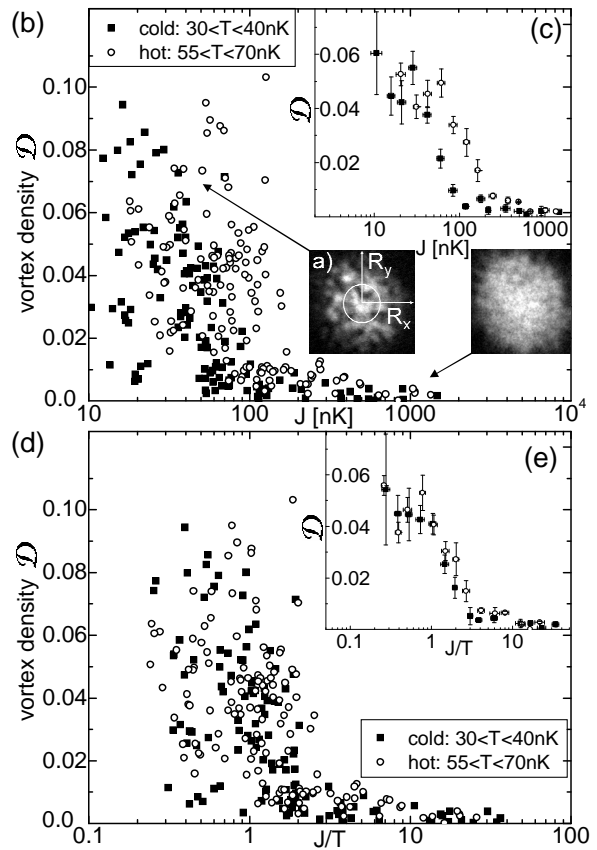,width=0.9\linewidth,clip=} \caption {\label{Fig2}Quantitative study of vortices. The areal density of
vortices is quantified by the plotted $\mathcal{D}$ defined in the text. $\mathcal{D}$ is extracted only from the central $11\%$ of the
condensate region [circle in inset (a)] to minimize effects of spatial inhomogeneity. (b) $\mathcal{D}$ vs $J$ for two datasets with distinct
``cold'' and ``hot'' temperatures. Each point represents one experimental cycle. The increase in $\mathcal{D}$ with decreasing
$J\lesssim100\,nK$ signals the spontaneous appearance of vortices, while the ``background'' $\mathcal{D}\lesssim0.01$ for $J\gtrsim200\,nK$ is
not associated with vortices. Vortices clearly proliferate at larger $J$ for the ``hot'' data, indicating thermal activation as the underlying
mechanism. The large scatter in $\mathcal{D}$ at low $J$ is due to shot noise on the small average number of vortices in the central condensate
region. (c) same data as in (b), but averaged within bins of size $\Delta[\log(J)]=0.15$. Error bars of $\mathcal{D}$ are standard errors. (d)
same data as (b), but plotted vs $J/T$. ``Cold'' and ``hot'' datasets almost overlap on what appears to be a universal vortex activation curve,
as confirmed by averaging [inset (e)], clearly revealing the underlying competition of $J$ and $T$.}
\end{figure}
\par
To investigate the thermal nature of phase fluctuations, we study vortex activation while varying $J$ at different temperatures. For a
quantitative study, accurate parameter estimates are required. The Josephson-coupling energy $J$ is obtained from 3D numerical simulations of
the Gross-Pitaevskii equation (GPE) for the central double-well system \cite{Bergeman,Oberthaler} [Fig.\ref{Fig1}(b)], self-consistently
including mean-field interactions of both condensed and uncondensed atoms \cite{Stamper-Kurn}. A useful approximation for $J$ in our experiments
is \cite{Javeragenote}: $J(V_{OL},N_{well},T)\approx N_{well}\times0.315\,nK\exp[N_{well}/3950-V_{OL}/244Hz](1+0.59\,T/100nK)$. The finite-$T$
correction to $J$ arises from both the lifting-up of the BEC's chemical potential and the axial compression by the thermal cloud's repulsive
mean field, but does {\it not} take into account the effects of phase fluctuations on $J$ (in condensed-matter language, we calculate the {\it
bare} $J$). $N_{well}$ is determined by comparison of the experimentally measured $R_z$, to $R_z(V_{OL},N_{well},T)$ obtained from GPE
simulations. Both experimental and simulated $R_z$ are obtained from a fit to the distribution of condensed and uncondensed atoms, to a
Thomas-Fermi profile plus mean-field-modified Bose function \cite{Stamper-Kurn}. In determination of all $J$ values, there is an overall
systematic multiplicative uncertainty $\Delta J/J=\,^{\times}_{\div}1.6$, dominated by uncertainties in the optical lattice modulation contrast,
the absolute intensity calibration, and magnification in the image used to determine $N_{well}$. In comparing $J$ for ``hot'' and ``cold''
clouds (see Fig. \ref{Fig2}) there is a relative systematic error of $15\%$ associated with image fitting and theory uncertainties in the
thermal-cloud mean-field correction to $J$.
\par
The qualitative results of our work are consistent whether we use an automated vortex-counting routine or count vortices by hand, but the former
shows signs of saturation error at high vortex density, and the latter is vulnerable to subjective bias. As a robust vortex-density surrogate we
therefore use the ``roughness'' $\mathcal{D}$ of the condensate image caused by the vortex cores. Precisely, we define $\mathcal{D}$ as the
normalized variance of the measured column density profile from a fit to a smooth finite-$T$ Bose profile \cite{Stamper-Kurn}, with a small
constant offset subtracted to account e.g. for imaging noise. To limit spatial inhomogeneity in $J$, caused by spatially varying condensate
density and optical lattice intensity, to $<10\%$, $\mathcal{D}$ is extracted only from the central $11\%$ of the condensate area which contains
20 lattice sites [Fig. \ref{Fig2}(a)]. Comparison to automated vortex-counts shows that $\mathcal{D}$ is roughly linear in the observed number
of vortices, irrespective of the sign of their circulation, with a sensitivity of $\approx0.01/$vortex.

\begin{figure}
\epsfig{figure=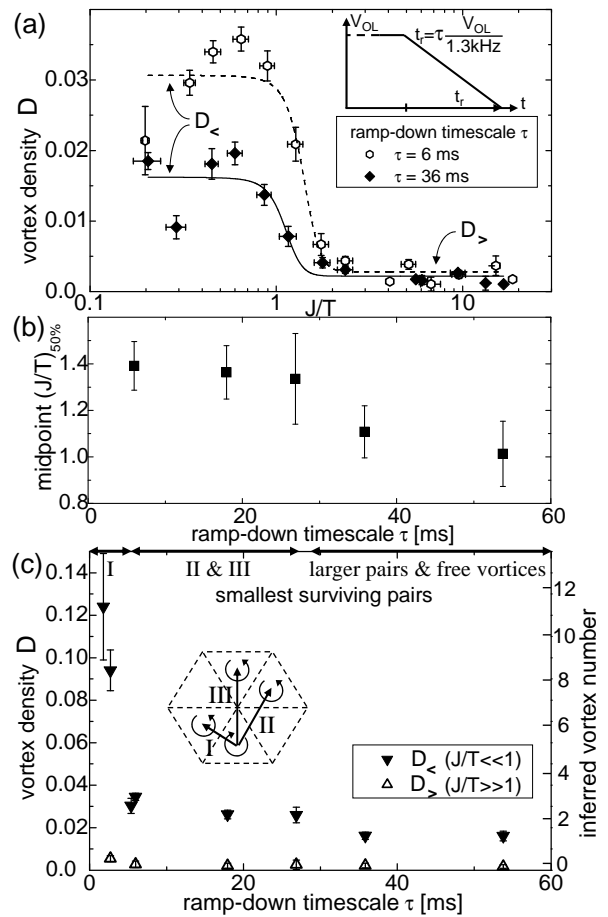,width=0.9\linewidth,clip=} \caption {\label{Fig3}(a) Vortex density $\mathcal{D}$ probed at different
optical lattice ramp-down timescales $\tau$. A slow ramp provides time for tightly bound vortex-antivortex pairs to annihilate, allowing
selective counting of loosely bound or free vortices only, whereas a fast ramp probes both free and tightly bound vortices. A fit to the vortex
activation curve determines its midpoint $(J/T)_{50\%}$, its $27\%-73\%$ width $\Delta(J/T)_{27-73}$, and the limiting values $\mathcal{D}_<$
($\mathcal{D}_>$) well below (above) $(J/T)_{50\%}$. (b) A downshift in $(J/T)_{50\%}$ is seen for slow ramps, consistent with the occurrence of
loosely bound or free vortices at lower $J/T$ only. (c) Mapping between ramp-down timescale $\tau$ and estimated size of the smallest pairs
surviving the ramp (upper axis). The difference $\mathcal{D}_< - \mathcal{D}_>$ measures the number of vortices surviving the ramp (right axis).
Comparison to simulated vortex distributions yields a size estimate of the smallest surviving pairs (upper axis). Inset: smallest possible pair
sizes in a hexagonal array, I: $d/\sqrt{3}$, II: $d$, III: $2d/\sqrt{3}$.}
\end{figure}
\par
Figure \ref{Fig2} shows results of our quantitative study. In Fig. \ref{Fig2}(b), we plot $\mathcal{D}$ vs $J$ for two datasets with distinct
temperatures. At large $J\gtrsim200\,nK$ a background $\mathcal{D}\lesssim0.01$ is observed, that is not associated with vortices, but due to
residual density ripples remaining after the optical lattice ramp-down. Vortex proliferation, signaled by a rise of $\mathcal{D}$ above
$\approx0.01$, occurs around $J\approx100\,nK$ for ``hot'' BECs and at a distinctly lower $J\approx50\,nK$ for ``cold'' BECs [confirmed by the
averaged data shown in Fig. \ref{Fig2}(c)], indicating thermal activation as the vortex formation mechanism. Plotting the same data vs $J/T$ in
Fig. \ref{Fig2}(d) shows collapse onto a universal vortex activation curve, providing strong evidence for thermal activation. A slight residual
difference becomes visible in the averaged ``cold'' vs ``hot'' data [Fig.\ref{Fig2}(e)], perhaps because of systematic differences in our
determination of $J$ at different temperatures.
\par
The vortex density $\mathcal{D}$ by itself provides no distinction between {\it bound} vortex-antivortex {\it pairs} and {\it free} vortices. In
the following we exploit the flexibility of optical potentials to distinguish free or loosely bound vortices from tightly bound
vortex-antivortex pairs. A ``slow'' optical lattice ramp-down allows time for tightly bound pairs to annihilate before they can be imaged. By
slowing down the ramp-down duration $\tau$ [inset of Fig. \ref{Fig3} (a)], we therefore selectively probe vortices at increasing spatial scales.
This represents an attempt to approach the ``true'' BKT vortex unbinding crossover that is complementary to transport measurements employed so
successfully in superconductive and liquid Helium systems.
\par
Figure \ref{Fig3}(a) shows vortex activation curves, probed with two different ramp-down times. Two points are worth noticing: First, a slow
ramp compared to a fast one shows a reduction of the vortex density $\mathcal{D}_<$ in arrays with fully randomized phases at low $J/T$. The
difference directly shows the fraction of tightly bound pairs that annihilate on the long ramp. Second, a slower ramp shows vortex activation at
lower $(J/T)_{50\%}$, confirming that free or very loosely bound vortices occur only at higher T (lower $J$). Specifically, the data clearly
show a range around $J/T\approx1.4$ where only tightly bound pairs exist. Figure \ref{Fig3}(b) quantitatively shows the shift of $(J/T)_{50\%}$
from 1.4 to 1.0 with slower ramp time. We can make a crude mapping of the experimental ramp-down time-scale to theoretically more accessible
vortex-antivortex pair sizes as follows: In Fig. \ref{Fig3}(c), we see the decrease of the saturated (low-$J/T$) vortex density $\mathcal{D}_< $
with increasing ramp timescale $\tau$. The right axis shows the inferred number of vortices that survived the ramp. We compare this number of
surviving vortices to simulations \cite{Simulation} of a 20-site hexagonal array with random phases. In these simulations we find, on average, a
total of 10 vortices, 6 of which occur in nearest-neighbor vortex-antivortex pairs [configuration I in Fig. \ref{Fig3}(c)], 1.7 (0.4) occur in
configuration II (III) respectively, and 1.9 occur in larger pairs or as free vortices. Experimentally $\approx11$ vortices are observed for the
fastest ramps, in good agreement with the expected ${\it total}$ number of vortices. For just somewhat slower ramps of $\tau\approx5\,ms$, only
3 vortices survive, consistent with only vortices in configuration II $\&$ III or larger remaining (indicated in Fig. \ref{Fig3}, top axis)
\cite{Config I decay}. For $\tau\gtrsim 30\,ms$ ramps less than 2 vortices remain, according to our simulations spaced by more than
$2d/\sqrt{3}$. Thus we infer that ramps of $\tau\approx30\,ms$ or longer allow time for bound pairs of spacing $\lesssim 2d/\sqrt{3}$ to decay
before we observe them. The downward shift of $(J/T)_{50\%}$ in Fig. \ref{Fig3}(b) thus tells us that loosely bound pairs of size larger than
$2d/\sqrt{3}$, or indeed free vortices, do not appear in quantity until $J/T\leq1.0$, whereas more tightly bound vortex pairs appear in large
number already for $J/T\leq1.4$.
\par
A further interesting observation concerns the width of the vortex activation curve. The relative width, determined from fits to data such as
the ones shown in Fig. \ref{Fig3}(a), is $\Delta(J/T)_{27-73}/(J/T)_{50\%}\approx0.3$, independent of ramp-down duration. This width is neither
as broad as in a double-well system \cite{Stringari,Oberthaler}, where the coherence factor rises over a range
$\Delta(J/T)_{27-73}/(J/T)_{50\%}\approx 1.4$, nor as broad as expected from our simulations \cite{Simulation} of an array of uncoupled phases,
each fluctuating independently with $\Delta\phi_{RMS}=\sqrt{T/J}$, for which we find $\Delta(J/T)_{27-73}/(J/T)_{50\%}\approx 0.85$. Presumably
collective effects in the highly multiply connected lattice narrow the curve. On the other hand, the width is 3 times larger than the limit due
to spatial inhomogeneity in $J$, suggesting contributions to the width due to finite-size effects or perhaps revealing the intrinsically smooth
behavior of vortex activation in the BKT regime.
\par
In conclusion, we have probed vortex proliferation in the BKT regime on a 2D lattice of Josephson-coupled BECs. Allowing variable time for
vortex-antivortex pair annihilation before probing the system provides a time-to-length mapping, which reveals information on the size of pairs
with varying $J/T$. We acknowledge illuminating conversations with Leo Radzihovsky and Victor Gurarie. This work was funded by NSF and NIST.


\begin{thebibliography}{28}
\bibitem{RadialEffects}
To avoid radial flows during $V_{OL}$ ramp-up, $R_{TF}$ is kept constant by balancing the lattice-enhanced mean field pressure with radial
confinement due to the optical lattice envelope, by the choice of a $67\mu m$ $1/e^2$ intensity waist. Axial (z) confinement is due to the
magnetic trap alone.
\bibitem{Axialphase}
In the axial condensate region between $z=-R_z/3,+R_z/3$, where according to our 3D GPE simulations $85\%$ of the tunnel current is localized
and hence the relative phase is measured, axial phase fluctuations \cite{Walraven} vary between $\approx 600\,mrad$ (``cold'' data in Fig.
\ref{Fig2}) and $800\,mrad$ (``hot'') in the regime $J/T\approx1$.
\bibitem{rampdown}
Within a dataset, the ramp-down rate is kept fixed, $t_r=\tau\times V_{OL}/1.3\,kHz$, $\tau=18\,ms$ if not otherwise indicated.
\bibitem{Javeragenote}
$J$ is averaged over junctions within the central $11\%$ of the array area, from which all quantitative experimental results are extracted.
\bibitem{Simulation}
Following \cite{Anderson}, in simulations we count a vortex if all three phase differences in an elemental triangle of junctions are
$\in(0,\pi)$, or if all are $\in(-\pi,0)$.
\bibitem{Config I decay}
The very short annihilation time of configuration I pairs [ $\tau < 5\,ms$ in Fig. \ref{Fig3}(c) ] is not unexpected: their spacing
$d/\sqrt{3}=2.8\,\mu m$ is comparable to the diameter of a vortex core in the bulk condensate (after $V_{OL}$ rampdown).
\end{thebibliography}

\end{document}